\documentstyle[prl,aps]{revtex}

\begin{document}

\tighten

\pagestyle{myheadings}
\markright{To appear in Physical Review Letters}

\draft

\title{
\rightline{\small{\sl To appear in Physical Review Letters}}
	First-order symmetric-hyperbolic Einstein equations with
       	arbitrary fixed gauge
       }

\author{Simonetta Frittelli\thanks{e-mail: simo@artemis.phyast.pitt.edu}}
\address{
	 Department of Physics and Astronomy, University of Pittsburgh,
	 Pittsburgh, PA 15260.
	}
\author{Oscar A. Reula\thanks{e-mail: reula@fis.uncor.edu}
		      \thanks{Member of CONICET}}
\address{
	 FAMAF, Universidad Nacional de C\'{o}rdoba, 
	 5000 C\'{o}rdoba, Argentina.
	}

\date{April 29, 1996}

\maketitle

\begin{abstract}

	We find a one-parameter family of variables which recast the
	3+1 Einstein equations into first-order symmetric-hyperbolic
	form for any fixed choice of gauge.  Hyperbolicity
	considerations lead us to a redefinition of the lapse in terms
	of an arbitrary factor times a power of the determinant of the
	3-metric; under certain assumptions, the exponent can be chosen
	arbitrarily, but positive, with no implication of gauge-fixing.

\end{abstract}

The issue of setting up a well-posed initial-value formulation for
general relativity has been studied with the help of varied strategies,
including special gauges and higher-order
formulations~\cite{choquet-york80}.  Recently, a renewed
interest~\cite{recent} in the problem has arisen, in connection with
the numerical evolution of the Einstein equations away from an initial
hypersurface.  Although the relevance of a manifestly hyperbolic
formulation to the numerical integration of the Einstein equations is
not yet clear, it is believed that a code tailored in a hyperbolic
formulation would share properties of the exact system; namely, it
would guarantee uniqueness and stability of solutions evolved from
proper initial data.  However, technical issues, associated with the
discretization of the equations and the precision of the approximation,
which may concern numerical stability, are not necessarily ruled out by
a pure theoretical hyperbolic development.

In regards to the manifest hyperbolicity of the Einstein equations, the
relevance of gauge choices has long been a question open to
consideration. The gauge freedom of 3+1 general relativity is embodied
by the lapse function and shift vector, which are completely arbitrary
since their evolution is not determined by the theory.  In general, a
theory expressed in terms of equations on fields which admit gauge
freedom may not admit a well-posed formulation unless improper gauge
choices are ruled out, or the true gauge-invariant variables of the
theory are found. A typical example is Maxwell's theory; on the one
hand, it admits a hyperbolic formulation at fixed gauge and in terms of
gauge invariant variables as well; on the other hand, anomalous gauges
can be found for which the resulting system does not have a well-posed
initial-value formulation.

Our intention is to give an explicit argument to rewrite the 3+1
Einstein equations into a manifestly well-posed form, without the need
of resorting to a choice of gauge. It has certainly been known that
general relativity in special gauges can be set in symmetric hyperbolic
form~\cite{earliersym}.  Furthermore, it has recently been shown that,
for certain special first-order variables, general relativity admits a
symmetric hyperbolic formulation for arbitrary fixed
gauge~\cite{frittelli-reula94,hypred}.  Here we extend the existing
results by showing that, under certain assumptions, there is a
one-parameter family of new first-order variables for general
relativity which satisfy first-order symmetric hyperbolic evolution for
arbitrary but fixed choice of gauge.

Several concepts of hyperbolicity (e.g., strict, strong) can assert the
well-posedness of a system of PDE's.  Among all different concepts,
symmetric hyperbolicity is especially appealing, for the reason that
most interesting physical systems admit a formulation of this
type~\cite{geroch95}.  Symmetric hyperbolicity is based on the symmetry
properties of the differential operator~\cite{courant-hilbert-john};
therefore, multiple eigenvalues, which usually occur due to the
presence of symmetries, play no role, as opposed to the case of other
types of hyperbolicity.  The reason for the well-posedness in the
symmetric-hyperbolic case is that the energy norm (an integral
expression in terms of the fields) at later times can still be seen to
be bounded by the norm at the initial time, because of cancellation of
terms under integration by parts.  The symmetry of the differential
operator in the evolution equations guarantees the cancellation.

In the following, we set up the problem of general relativity in the
3+1 formulation due to Arnowitt, Deser and Misner in a non-canonical
(though widely used) choice of variables; i.e., the intrinsic metric
and the extrinsic curvature of the spatial hypersurfaces.  We then
redefine the variables in order to reduce the system to first order;
the redefined variables depend on a set of parameters to be fixed by
hyperbolicity considerations.  Finally,  we use the argument of the
cancellation of terms under integration by parts in the energy norm to
determine the parameters.  In the process, we find that the lapse
function must be redefined in terms of the determinant of the 3-metric,
without any loss of gauge freedom.

In order to fix notation, we summarize some necessary points of the 3+1
formulation.  The 3+1 splitting of the fully 4-dimensional formalism
consists of a spacelike foliation by the level surfaces of a function
$t(x^a)$. The unit normal form is $n_a=-N\nabla_a t$, where $N$
is the lapse function. The unit normal vector is given by
$n^a=-\frac{1}{N}(t^a-N^a)$, where $N^a$ is the shift
vector.  The metric $g^{ab}$ induces a 3-metric on the spatial
surfaces, by $h^{ab}= g^{ab} + n^a n^b$.  In a coordinate
system $\{ x^0,x^i\}$, $i=1,2,3$, adapted to the surfaces (such that
$t=x^0$), the induced metric $h^{ab}$ reduces to $h^{ij}$.  The
extrinsic curvature of the 3-surfaces is defined by $K^{ab}\equiv
\frac12\pounds_{\!n} h^{ab}$, and is also a 3-tensor.

The equations for the evolution of the intrinsic contravariant metric
$h^{ij}$ and the extrinsic curvature $K^{ij}$ can be taken as (from
\cite{york79}, with the notation of Chapter 10 and Appendix E of
\cite{wald84})
   \begin{eqnarray}
	\dot{h}^{ij} 				&
						     =
						&
	2 N   K^{ij} - D^i N^j
		     - D^j N^i					
							\label{hdot}	\\
	\dot{K}^{ij} 				&
						     =
						&
	  N \Big(
	          	R^{ij} 		- 
		      2 K^{ik}{K^j}_k	+
			K^{ij} K			\nonumber\\
	& &				-
		 \kappa  \big(	      S^{ij}			-
			      \frac12 h^{ij} ( S  - \rho )
			 \big)
	    \Big)			-
			       D^iD^j N			\nonumber\\	
	& &					+
	  		N^ k   D_k K^{ij}	-
	  		K^{ik} D_k N^j		-
			K^{jk} D_k N^i			\label{kdot}
  \end{eqnarray}
The notation $(\,\dot{\;}\,)$ stands for $\pounds_{t^a}$ or simply
$\partial/\partial t$.  Indices $i,j,k,\ldots$ are raised and
lowered with the 3-metric $h^{ij}$; the operator $D_i$ is the covariant
derivative with respect to the 3-metric $h_{ij}$~\cite{wald84}. For any
3-tensor $U^{ij}$, the notation $U$ stands for its trace with respect
to the 3-metric: $U \equiv {U^k}_k$.  The matter tensor $S^{ij}$ is the
projection of the 4-dimensional stress-energy tensor $T^{ab}$ into
the spatial hypersurfaces, and $\rho$ is the projection of $T^{ab}$
in the direction normal to the surfaces.  The results derived in this
Letter do not depend strongly on the particular matter source, but hold
for any sources that admit a first-order symmetric-hyperbolic
formulation on their own.

Equations (\ref{hdot}) and (\ref{kdot}) for the fields $(h^{ij},K^{ij})$ are supplemented by the constraints
  \begin{eqnarray}
       {\cal C}\;\, & := &\; \frac12(
			      R + K^2 - K_{ij}K^{ij}
				     )             - \kappa \rho  = 0,   
							\label{scalar}\\
       {\cal C}^i   & := & \;D_j K^{ij} - D^i K    - \kappa  J^i  = 0,    
							\label{vector}
  \end{eqnarray}
where $  J^i $ is the mixed projection of $T^{ab}$ onto the
hypersurface and the normal.  If $\cal C$ and ${\cal C}^i$ can be shown
to be conserved as a consequence of (\ref{hdot}) and (\ref{kdot}), then
the constraints only need to be imposed on an initial hypersurface.
This will be our point of view in the following.

\paragraph*{Introduction of the parameters.}  In order to show that
there exists a one-parameter family of variables for which general
relativity takes a first-order symmetric-hyperbolic form, we first
introduce a set of four parameters, $\alpha,\beta,\gamma$ and
$\epsilon$; we eventually require the parameters to satisfy a set of
three algebraic conditions that guarantee the hyperbolicity.

	Two parameters, $\alpha$ and $\beta$, are used to redefine
	variables as follows
     \begin{eqnarray}
	{M^{ij}}_k & \equiv & \frac12 (	
			        h^{ij}\!,_k 	     + \;
		       \alpha\; h^{ij} 
				h_{rs} h^{rs}\!,_k   ),	 \label{mabc}
									\\
	P^{ij}     & \equiv & K^{ij} +\; \beta \; h^{ij} K.
							 \label{pab}
    \end{eqnarray}
	The definition (\ref{mabc}) reduces the Einstein equations to
	first order.  Note that the variable ${M^{ij}}_k$ represents
	the spatial derivative of the densitized 3-metric,
       	$	{M^{ij}}_k = 		\frac12  h^{\alpha}
				\big(
			             	         h^{-\alpha}
		    			         h^{ij}
				\big),_k
	$, where $h$ is the determinant of $h_{ij}$.
	Equations (\ref{mabc}) and (\ref{pab}) can be inverted into
    \begin{eqnarray}
	h^{ij}\!,_k & \equiv & 2 \; (\,	
       {M^{ij}}_k 	       - \; \frac{\alpha}{3\alpha \!+\! 1}
        h^{ij}
	M_k  \, ),	 \label{mabc:inv}
									\\
	K^{ij}     & \equiv & 
	P^{ij} 		      -\; \frac{\beta }{3\beta  \!+\! 1}
	h^{ij} P,
							 \label{pab:inv}
    \end{eqnarray}
	with the notation $
	M_k	      \equiv h_{ij} {M^{ij}}_k	$.

	 A third parameter, $\gamma$, is introduced in the evolution
	equations to allow for a combination of (\ref{hdot}) and
	(\ref{kdot}) with the constraints (\ref{scalar}) and
	(\ref{vector}). In this way,  the principal part of the
	evolution equations (\ref{hdot}) and (\ref{kdot}) can be
	modified. The constraints (\ref{scalar}) and (\ref{vector})
	will be assumed to be conserved by the resulting equations.
	Since $\gamma$ plays a crucial role in the hyperbolicity of the
	system, in the following we point out its exact place in the
	evolution equations.

	The evolution equation for ${M^{ij}}_k$ can be obtained by,
	first, taking a space derivative $\partial/\partial x^k$ of
	Eq.(\ref{hdot}), and then tracing and combining the resultant
	equation according to the definition (\ref{mabc}).  We also add
	the vector constraint ${\cal C}^i$ with an appropriate
	--uniquely determined-- factor, obtaining the following:
  \begin{eqnarray}
	\dot{M}^{ij}{}_k & = &
			   \frac12 \Big(
			   \big(
				  \dot{h}^{ij}	
			   \big),_k		  +	\alpha
			   \big(
				  \dot{h}^{ij}
				       h_{rs}	  +
				       h^{ij}
				  \dot{h}_{rs}
			   \big)       h^{rs},_k  	\nonumber\\
		& & 				  +	\alpha
				       h^{ij}
				       h_{rs}
			   \big(
				  \dot{h}^{rs}
			   \big),_k \Big)		  
						  -\;
			   	N\delta^{(i}_k{\cal C}^{j)}\;. \label{evolm}
  \end{eqnarray}
	If Eq.(\ref{hdot}) is used in the right-hand side of
	(\ref{evolm}) to eliminate time derivatives in favor of space
	derivatives of the new fields, the right-hand side becomes a
	combination of the fields $h^{ij},{M^{ij}}_k,P^{ij}$, lapse,
	shift and sources; first-space derivatives of
	${M^{ij}}_k,P^{ij}$, lapse and shift; and second-space
	derivatives of the shift.  This equation is shown explicitly
	below [Eq. (\ref{pp.evolm})], correct to principal terms.

	The evolution equation for $P^{ij}$ is obtained directly from
	Eq.(\ref{kdot}), by the appropriate combination with its trace,
	as prescribed by the definition (\ref{pab}).  We also add the 
	scalar constraint ${\cal C}$ with a suitable factor:
	\begin{eqnarray}
	   \dot{P}^{ij} & = &  
			  \dot{K}^{ij}		+ \beta \Big(
			  \dot{h}^{ij} 
			       K		+
			       h^{ij}
			  \dot{h}_{rs}
			       K^{rs}		+
			       h^{ij}
			       h_{rs}
			  \dot{K}^{rs}			\Big) \nonumber\\
		& &				+
	             2N\gamma h^{ij} {\cal C}.	\label{evolp}
	\end{eqnarray}
	When Eqs. (\ref{hdot}) and (\ref{kdot}) are substituted
	appropriately in (\ref{evolp}), the right-hand side becomes a
	combination of the fields $h^{ij},{M^{ij}}_k,P^{ij}$, lapse,
	shift and sources; first-space derivatives of
	${M^{ij}}_k,P^{ij}$, lapse and shift; and second-space
	derivatives of lapse.  This equation is shown explicitly below
	[Eq. (\ref{pp.evolp})], correct to principal terms.

        The fourth parameter, $\epsilon$, is introduced in order to 
	redefine the lapse $N$ by
	 \begin{equation}
	      N      \equiv  h^{-(3\alpha + 1)\epsilon/2} Q\;\;,
         					\label{lapseQ}
	 \end{equation}
        for an arbitrary function $Q$. The lapse is thus redefined
	without loss of generality; the gauge freedom is transferred to
	$Q$, and the parameter $\epsilon$ remains to be specified.
	Notice that
	 \begin{equation}
	\frac{N,_k}
	      N      \equiv \epsilon            M_k  	+
					(\ln Q ),_k     \;\;.
		 					\label{lapse}
 	 \end{equation}
	Since second derivatives of the lapse appear in (\ref{evolp}),
	this redefinition allows for a modification of the principal
	terms in (\ref{evolp}).

\paragraph*{Hyperbolicity imposed on the system.}

We define the energy norm of the system at time $t$ as
  \begin{equation}
   	E(t) 	= \frac12
		  \int_{\Sigma} h^{ij}    h_{ij} 	+ 
				P^{ij}    P_{ij} 	+ 
			       {M^{ij}}_k{M_{ij}}^k
  \end{equation}
where the integration is performed on the surface $\Sigma$ defined by
$t=const$. The spatial symmetry of the system is guaranteed if the
principal terms in the time derivative of the energy~\cite{lax55} can
be combined into total divergences, since in this case their
contribution to the energy estimates would vanish.

The time-derivative of the energy, correct to principal terms, is
  \begin{equation}
    \dot{E}(t)  =    \int_{\Sigma} \dot{h}^{ij}      h_{ij} 	+ 
				   \dot{P}^{ij}      P_{ij} 	+ 
			           \dot{M}^{ij}{}_k{M_{ij}}^k\;\;.
							\label{edot}	
  \end{equation} 
The evolution equations (\ref{hdot}), (\ref{evolm}) and (\ref{evolp})
can be used to trade time derivatives for space derivatives in
(\ref{edot}). If the principal terms can be eliminated under
integration by parts, then the system becomes hyperbolic.

In the following we write the principal terms of the evolution 
equations and find the conditions that are necessary
to symmetrize the system.

The Ricci tensor $R_{ij}$ is needed in terms of the new fields. Recall
\[ 	R_{ij} = \Gamma^k_{ij,k}  \;-\; 
		 \Gamma^k_{ki,j}  \;+\;
		 		        \Gamma^k_{ij} \Gamma^l_{kl} \;-\;
				        \Gamma^k_{lj} \Gamma^l_{ki}\;\;,
\]
with
$
	\Gamma^k_{ij} =   -\;\frac12 h_{il} h^{kl}\!,_j
			\;-\;\frac12 h_{jl} h^{kl}\!,_i
			\;+\;\frac12 h^{kl} 
				     h_{ir}
				     h_{js} h^{rs}\!,_l.
$
In terms of ${M^{ij}}_k$, the connection $\Gamma^k_{ij}$ takes the form
  \begin{eqnarray}
	\Gamma^k_{ij} & = &  -\; 2 h_{l(i} {M^{kl}}_{j)}
			\;+\;   h^{kl} 
				h_{ir}
				h_{js}  {M^{rs}}_l
			\;+\; \frac{2\alpha}{3\alpha \!+\! 1}
				\delta^k_{(i} M^{}_{j)}	\nonumber\\
		& &	\;-\; \frac{ \alpha}{3\alpha \!+\! 1}
 				h_{ij} 
				h^{kl}  M_l.
  \end{eqnarray}
The principal part of (\ref{evolp}) is then
  \begin{eqnarray}
	\dot{P}^{ij} & = & N^k           P^{ij},_k
			\;+\;
			   N \Big(
			  	h^{kl}  {M^{ij}}_k,_l
			\;-\; 2 h^{l(i} {M^{j)k}}_l,_k		\nonumber\\
	& &		\;+\;     \frac{2\alpha \!+\! 1}
				       {3\alpha \!+\! 1}
				h^{ik}
				h^{jl}   M_k,_l
			\;+\;    \frac{ 2\beta(\alpha\!+\!1)\!-\!\alpha}
				      {3\alpha \!+\! 1}
 				h^{ij} 
				h^{kl}   M_k,_l			\nonumber\\
	& &		\;-\;           2\beta
				h^{ij}  {M^{kl}}_k,_l	
			     \Big)
			\;-\;
				h^{ik}
				h^{jl}   N,_{kl} - \beta 
				h^{ij} 
				h^{kl}   N,_{kl}		\nonumber\\
	& &		\;+\; 		2N\gamma
				h^{ij}
			     \Big(
			  -		{M^{kl}}_k,_l
			\;+\;  
			      \frac{ \alpha \!+\! 1}{3\alpha \!+\! 1}
				h^{kl}   M_k,_l
			     \Big)\;\;.
							\label{pp.evolp}
  \end{eqnarray}
The principal terms of (\ref{evolm}) are the following:
  \begin{eqnarray}
	\dot{M}^{ij}{}_k & = & 
			   N^l         {M^{ij}}_k,_l	
				\;+\;
			   N \Big(
				   	P^{ij},_k 
				\;+\;	\frac{\alpha\!-\!\beta}
					     {3\beta\!+\!1}
					h^{ij}
					P,_k
			     \Big)		\nonumber\\
		 & &		\;-\;
					2 N \delta^{(i}_k
					P^{j)l},_l
				\;+\;   
					2 N \frac{\beta\!+\!1}
					         {3\beta\!+\!1}
					    \delta^{(i}_k
					h^{j)l}
					P,_l	\;\;.	\label{pp.evolm}
  \end{eqnarray}

In view of (\ref{pp.evolp}) and (\ref{pp.evolm}), the cancellation
under integration by parts in (\ref{edot}) takes place if the following
algebraic conditions are imposed on the parameters
$\alpha,\beta,\gamma$ and $\epsilon$:
 \begin{mathletters}
   							\label{allequations}
  \begin{eqnarray} 
	\frac{2\alpha+1}
	     {3\alpha+1} -        \epsilon   =  0,	\label{equationa}	\\	
	\frac{ \beta +1}
	     {3\beta +1} + \beta + \gamma    =  0,	\label{equationb}	\\	
	\frac{2(\beta+\gamma)(\alpha+1)-\alpha}
	     {3\alpha+1} - 
	\frac{\alpha-\beta}
	     {3\beta+1}  - \beta  \epsilon   =  0.	\label{equationc}
  \end{eqnarray}
 \end{mathletters}
Condition (\ref{equationa}) has the effect of the cancellation of the
fourth and seventh terms in (\ref{pp.evolp}), even before their
contribution to the energy is considered.  This is done in this way,
because the fourth term in (\ref{pp.evolp}) has no symmetric counterpart
in (\ref{pp.evolm}) with respect to its contribution to the energy,
and, therefore, needs to be eliminated from the system. Condition
(\ref{equationb}) guarantees the cancellation, under integration by
parts, of the sixth and ninth terms in (\ref{pp.evolp}), together with
their symmetric counterpart, i.e. the fifth term in
(\ref{pp.evolm}).  Lastly, condition (\ref{equationc}) guarantees the
symmetry of the fifth, eighth and tenth terms in (\ref{pp.evolp}) with
the third term in (\ref{pp.evolm}), which subsequently make no
contribution to $\dot{E}$.

With the assumptions that $Q>0$, that $h^{ij}$ is positive definite,
that the algebraic conditions (\ref{allequations}) are met by the four
parameters $\alpha,\beta,\gamma$ and $\epsilon$, and that the
constraints $\cal C$ and ${\cal C}^i$ are conserved, the fields
$(h^{ij}, {M^{ij}}_k,P^{ij})$ satisfy a symmetric hyperbolic system of
PDE's, namely Eqs.  (\ref{hdot}), (\ref{evolm}) and (\ref{evolp}), with
the initial data constrained by (\ref{scalar}), (\ref{vector}) and
(\ref{mabc}).

Notice that the conditions (\ref{allequations}) leave free one of the
four parameters.  Any one of the parameters can be chosen freely,
within a real range that allows for real values for the remaining three
parameters as solutions of (\ref{allequations}).  For instance, if
$\alpha$ is considered as the free parameter, then $\alpha$ can take
values in $ \big(-\infty,-1/2 \big) $, while $\beta$ must be chosen as
a root of the following quadratic equation:
  \begin{equation}
	3\beta^2+2\beta+\frac{(3\alpha+1)(\alpha+1) +1}
			     {(2\alpha + 1 )          } =0
  \end{equation}  

A most interesting choice of variables is $P^{ij} = K^{ij} - h^{ij}K$
(proportional to the canonical ADM momentum), or $\beta=-1$.  This
choice of $\beta$, in turn, fixes $\gamma=1$, $\alpha=-1$ and
$\epsilon=1/2$. The redefinition of lapse becomes $N=Q\sqrt{h}$, and
the number of terms in the principal parts in Eqs.  (\ref{pp.evolp})
and (\ref{pp.evolm}) reduces considerably. Regarding the propagation of
the constraints, for $\gamma=1$ it can be shown that the Bianchi
equations imply a homogeneous symmetric hyperbolic evolution system for
$\cal C$ and ${\cal C}^i$. It follows that the constraints are
conserved.  This case was explored earlier by the authors, and
has been found suitable for the development of a smooth newtonian
limit~\cite{frittelli-reula94} if certain gauge choices are imposed in
addition to the well-posed formulation. Most remarkably, for
$\gamma\neq 1$ the evolution of the constraints is not symmetric
hyperbolic nor strictly hyperbolic, and the validity of the assumption
of the conservation of the constraints must be studied carefully. The
details will soon appear elsewhere.

Eq. (\ref{equationa}) shows that the exponent of $\sqrt h$ in the
redefinition of the lapse (\ref{lapseQ}) is equal to $-(2\alpha+1)$,
being thus any positive real number, but never zero. Thus, it is not
possible to have a set of variables of the form
(\ref{mabc})-(\ref{pab}) with symmetric hyperbolic evolution without
relating the true lapse $N$ to the 3-metric.


The system (\ref{hdot}), (\ref{evolm}) and (\ref{evolp}) has a non
trivial set of characteristics~\cite{charac}. Using the notation
$\xi_a\equiv (v,\xi_i)$ where $\xi_i$ has unit norm with respect to
$h^{ij}$, it is immediate to see that $\xi_a$ is characteristic if $t^a
\xi_a=0$ (for any $\alpha$). Furthermore, by essentially the same
arguments as in ~\cite{hypred}, it can be shown that covectors $\xi_a$
satisfying either $\xi^a\xi_a=0$ or $n^a\xi_a=0$ are also
characteristic (for any $\alpha$).  There are no other characteristics
if $\alpha$ takes the value $-1$, as in~\cite{frittelli-reula94}.
Therefore, if $\alpha=-1$, the characteristics are null (with speeds
$N^i\xi_i\pm N$), or timelike and either tangent to $t^a$ (with zero
speed) or tangent to the normal direction $n^a$ (with speed
$N^i\xi_i$). However, if $\alpha\neq-1$ the system may have other
characteristics, {\em in addition to these\/}, with speeds that may
depend (non-trivially) on the choice of $\alpha$.  The details will
also appear elsewhere.

In this work the choice of $N^i$ and $Q$ is {\em arbitrary\/} but {\em
given\/}. The gauge must be specified in order to integrate  the
equations. On the other hand, the hyperbolicity of the system holds
independently of the choice of gauge.  The fact that gauge-fixing is
required is not troublesome; the variables themselves are not
gauge-invariant fields.

In order to avoid confusion, we point out that by fixing a gauge we
understand a non-dynamical specification of $N^i$ and $Q$ as {\em a
priori\/} known functions for all time, independent of the evolution of
the new fields.  In this way, $N^i$ and $Q$ act as known sources.  If
the gauge were specified dynamically (explicitly, or implicitly via an
equation) as a function of the new fields, then the principal part of
the evolution equations would be modified. The results proven here do
not guarantee the well-posed evolution of such a choice of gauge; in
fact, it is not hard to see that, in general, the well-posedness would
be hampered.

The system shown here shares, with other hyperbolic formulations, the
property of manifest first-order flux-conservative form, which makes it
suitable for the application of general numerical integration
techniques~\cite{lax73} (no numerical applications of this formalism
have been investigated as of now).  Aside from that, we find it very
appealing for its remarkable simplicity and clarity.

It is our pleasure to thank H. Friedrich for valuable suggestions and
criticisms.  O. A. Reula acknowledges support from CONICOR and SECYT
(Argentina). S. Frittelli acknowledges partial support from NSF.

\end{document}